\documentclass[journal]{IEEEtran} 
\usepackage{graphicx} 
\usepackage[center]{caption}
\usepackage{hyperref}
\usepackage{amsmath}
\usepackage{color}

\usepackage[normalem]{ulem} 
\newcommand{\stkout}[1]{\ifmmode\text{\sout{\ensuremath{#1}}}\else\sout{#1}\fi}

\begin{document}

\title{Assessing the importance of long-range correlations for deep-learning-based sleep staging}

\author{Tiezhi Wang and Nils Strodthoff% 
\thanks{The authors are with the Carl von Ossietzky Universität Oldenburg, Germany. Corresponding author: Nils Strodthoff (nils.strodthoff@uol.de)}
\thanks{}
}

\markboth{Proceedings of the WORKSHOP BIOSIGNAL 2024, February 28th\,-\, March 1st, 2024, Göttingen, Germany}%
{Kurths \MakeLowercase{\textit{et al.}}: Assessing the importance of long-range correlations for deep-learning-based sleep staging}

\maketitle

\begin{abstract}
This study aims to elucidate the significance of long-range correlations for deep-learning-based sleep staging. It is centered around S4Sleep(TS), a recently proposed model for automated sleep staging. This model utilizes electroencephalography (EEG) as raw time series input and relies on structured state space sequence (S4) models as essential model component. Although the model already surpasses state-of-the-art methods for a moderate number of 15 input epochs, recent literature results suggest potential benefits from incorporating very long correlations spanning hundreds of input epochs. In this submission, we explore the possibility of achieving further enhancements by systematically scaling up the model's input size, anticipating potential improvements in prediction accuracy. In contrast to findings in literature, our results demonstrate that augmenting the input size does not yield a significant enhancement in the performance of S4Sleep(TS). These findings, coupled with the distinctive ability of S4 models to capture long-range dependencies in time series data, cast doubt on the diagnostic relevance of very long-range interactions for sleep staging.	
\end{abstract}

\begin{IEEEkeywords}
Electroencephalography, Machine learning algorithms, Time series analysis
\end{IEEEkeywords}

\IEEEpeerreviewmaketitle

\section{Introduction}
Sleep disorders, affecting a considerable portion of the population, significantly impact health and well-being. Primary care frequently involves managing various sleep disorders. Accurate diagnosis and treatment of these conditions requires sleep staging, a process of classifying human into distinct stages using polysomnography (PSG). PSG integrates multiple sensor modalities, among which the electroencephalogram (EEG) is particularly relevant for most automatic sleep staging algorithms. Traditionally, sleep stages are annotated according to established guidelines \cite{iber2007aasm,hobson1969manual} in 30-second intervals known as epochs. However, the manual annotation is time-consuming and subject to a high inter-rater variability, underscoring the need for automated systems with performance comparable to human experts.

Recent advancements in deep learning have significantly improved automatic sleep staging. State-of-the-art models, including U-Sleep \cite{Perslev2021} and sequence-to-sequence architectures, have achieved expert-level accuracy. Recent works show a growing emphasis on encoder-predictor model architectures combining convolutional encoders with Long Short-Term Memory (LSTM) \cite{Phan2021} or transformer \cite{Brandmayr2021} predictor models. A notable recent innovation in model architecture are structured state space sequence (S4) models \cite{Gu2021EfficientlyML}. They consititute the essential building block of a recently proposed sleep staging model S4Sleep \cite{wang2023s4sleep} that utilizes S4 instead of LSTM or transformer layers, and demonstrated superior performance in sleep staging tasks, both for models operating on time series (S4Sleep(TS)) and on spectrograms (S4Sleep(Spec)). This model benefits from sub-epoch-level tokenization and uses only moderately long input sequences (15 epochs equivalent to 7.5 minutes).

%\added{In our research, we categorize input sequence sizes as either ``long'', denoting sequences longer than 5 minutes (equivalent to 10 epochs) but less than 50 minutes (equivalent to 100 epochs), or ``very long'', representing input sequences of 50 minutes or longer.} 
In sleep staging, as reviewed in a recent review \cite{fiorillo2019automated}, incorporating the long-term temporal context, in the sense of predicting sleep stages for $O(15)$ subsequent epochs at once, is an accepted paradigm \cite{phan2019seqsleepnet}. Still, it is known that neither LSTM nor transformer models are particularly effective in capturing very-long-term dependencies as demonstrated in their inability to solve long-range benchmark problems \cite{tay2021long,Gu2021EfficientlyML}. However, there exist long-range correlations of the heart rate in REM sleep phases \cite{PhysRevLett.85.3736}, which provides a physiological motivation for this study. In particular, it might be seen as a hint for a potential diagnostic relevance of such long-term dependencies. Recently, Phan et al \cite{phan2023lseqsleepnet} proposed L-SeqSleepNet arguing for the necessity of including long-range dependencies across hundreds of epochs. In this work, we revisit this hypothesis building on the ability of the S4 model to efficiently capture long-range interactions.

\section{Methods}

\subsection{Model}
\begin{figure}[!t]
	\centering
	\includegraphics[width=0.90\linewidth]{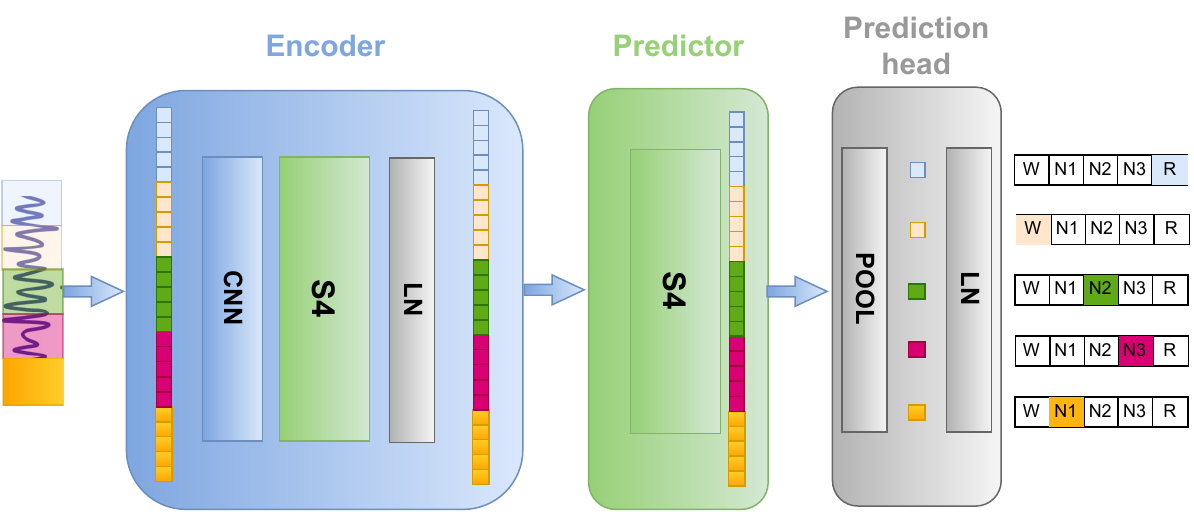}
	\caption{Schematic representation of the S4Sleep(TS) model, which is composed of a S4-model-based sub-epoch encoder and a S4-model-based predictor along with a local pooling and linear classifier as prediction head.}
	\label{arch}
\end{figure}

We base our work on the \textit{S4Sleep(TS)} model operating on raw time series, which demonstrated superior performance compared to \textit{S4Sleep(Spec)} using a single EEG channel as input. Through an extensive architecture search this architecture was identified as optimal architecture across a wide range of modular architecture components \cite{wang2023s4sleep} including transformer layers or recurrent layers. The S4Sleep(TS) model is an encoder-predictor model and comprises three main components: an epoch encoder, a predictor, and a prediction head. The \textit{epoch encoder} aggregates a fraction of 1/5th of an epoch, into a latent representation suitable for further processing. In the present model, the epoch encoder is comprised of two one-dimensional convolutions followed by a four-layer S4 model and an average pooling layer. This yields a latent representation summarizing 6s of the input signal. This heavily temporally downsampled representation is then processed by a \textit{predictor model}, which is in this case also given by a S4 model and outputs a processed representation of the same shape as the layer's input representation. Finally the predictor output is processed by a \textit{prediction head}, which is comprised of a local average pooling layer, which averages across 5 subsequent token to match the temporal resolution of the epoch-level annotation, followed by a linear output layer. We summarize the model architecture in Fig.~\ref{arch} and refer  to \cite{wang2023s4sleep} for further details.

\subsection{Datasets}
For this study we use the Sleep-EDF (SEDF) dataset, which is publicly available from PhysioNet and encompasses 197 recordings from 106 individuals \cite{kemp2000analysis, goldberger2000physiobank}. Despite its comparably small size, SEDF is widely used in the literature and results on SEDF were found to be largely consistent with those on the larger SHHS \cite{wang2023s4sleep} dataset. We resort to the former for computational reasons. Sleep stages in these recordings were manually annotated. Adhering to standard practices, we merge the stages N3 and N4 into a single category (N3) and exclude segments marked as MOVEMENT or UNKNOWN from the loss calculation and evaluation but retain the corresponding segments in the input sequences. For comparison with the literature, we base our experiments on a single EEG channel (Fpz-Cz) even though the proposed model is also able to achieve an improved performance by leveraging multi-channel input data \cite{wang2023s4sleep}. We use a data partitioning scheme that allocates 80\% of the recordings for training, 10\% for validation, and the remaining 10\% for testing, as described in \cite{wang2023s4sleep}. No additional preprocessing was applied to the raw time series data.

\subsection{Training procedure and performance evaluation}
To address the challenge of imbalanced label distribution typical for sleep stage classification tasks, we use the focal loss  as loss function. All experiments were conducted at a fixed batch size of 64, achieved through gradient accumulation, and a constant learning rate of 0.001 using the AdamW optimizer. During training and validation phases, the input sequences were segmented into consecutive, non-overlapping sections that matched the model's input size. For testing, the input sequences were divided into segments of equal length, albeit with a smaller stride corresponding to the length of a single epoch. This approach enabled multiple predictions for each segment, depending on its position in the input sequence. These predictions were then averaged at the output probability level for each segment, see \cite{wang2023s4sleep}.

Training models with very long input sizes is challenging from an optimizational perspective. This is a well-known phenomenon also in other fields, e.g., in natural language processing when training language models based on genomic data with input sizes of up to 1M tokens \cite{nguyen2023hyenadna}. One way of resolving this issue is to gradually increase the input size of the model while subsequently finetuning the model. In our case, the model was trained initially with an input sequence length of 10 epochs, equivalent to 5 minutes, for a total of 50 training epochs. Subsequently, the input sequence length was progressively doubled to 20, 40, 50, 100, and 200 epochs, with each extension undergoing an additional 10 epochs of finetuning. To avoid overfitting, we selected the best-performing training epoch based on the achieved macro $F_1$-score on the validation set. Following each training extension on the input sequence length, the model's performance was evaluated on the test set. 
 
Given the importance of accurately assessing the performance differences, an uncertainty estimate is essential. To address the uncertainty stemming from the finite and specific composition of the test set, we use empirical bootstrapping with 1,000 iterations on the test set. This method allowed us to calculate 95\% confidence intervals. We considered the performance differences between models to be statistically significant if the confidence intervals for the difference between two macro $F_1$-scores does not overlap with zero.

\section{Results}
The main results of our investigation are compiled in Tab.~\ref{table_example}, summarizing the model performance in dependence of the model's input size. In \cite{wang2023s4sleep}, it was already demonstrated that the proposed model is on par with state-of-the-art methods on SEDF and outperforms those on the larger SHHS dataset. However, here is it worth stressing that the best-performing literature result, L-SeqSleepNet\cite{phan2023lseqsleepnet}, was trained on very long input sequences of 200 epochs, while S4Sleep(TS) reaches a similar level of performance already at 10-15 epochs using raw waveforms as opposed to spectrograms as input representation. Tab.~\ref{table_example} shows no significant increase in performance upon increasing the model's input size. In fact, all results agree within error bars and there are no statistically significant differences among the models. Also note that the result for training a S4Sleep(TS) model directly on 200 input epochs is statistically significantly worse, which clearly demonstrates the need for a careful training schedule with a gradually increasing input size.

\begin{table}[!t]

\renewcommand{\arraystretch}{1.3}

\caption{Sleep staging performance results in dependence of the number of input epochs and in comparison to literature results.}
\label{table_example}
\centering

\begin{tabular}{p{2.9cm}p{0.4cm}p{0.6cm}p{1.1cm}p{1.8cm}}
\hline	
	
			 model & input epoch & feature	& base model & $\text{macro} F_1$ $(\uparrow)$ \\
		
\hline

		 S4Sleep(TS)\textsuperscript{c, f} & 10 & raw& S4  & 0.790$\pm$0.004 \\
		 S4Sleep(TS)\textsuperscript{c, f} & 20 & raw& S4  & 0.794$\pm$0.004 \\
		  S4Sleep(TS)\textsuperscript{c, f} & 40 & raw& S4  & 0.795$\pm$0.004 \\
		  S4Sleep(TS)\textsuperscript{c, f} & 50 & raw& S4  & 0.796$\pm$0.004 \\
		  S4Sleep(TS)\textsuperscript{c, f} & 100 & raw& S4  & 0.794$\pm$0.004 \\
		  S4Sleep(TS)\textsuperscript{c, f} & 200 & raw& S4  & 0.793$\pm$0.004 \\
		  S4Sleep(TS)*\textsuperscript{c, f} & 200 & raw& S4  & 0.723$\pm$0.004 \\
		 \hline
		 \textbf{S4Sleep(TS)\textsuperscript{c, f}}  \cite{wang2023s4sleep} & 15 & raw & S4 & \textbf{0.796$\pm$0.004}  \\
		L-SeqSleepNet\textsuperscript{a, g}\cite{phan2023lseqsleepnet} & 200 & spec& LSTM &  0.793$\pm$0.004  \\
		 SeqSleepNet\textsuperscript{a, d}\cite{phan2019seqsleepnet}\cite{phan2023lseqsleepnet} & 10 & spec & LSTM  &  0.786$\pm$0.002  \\ 
		 
		 SleepTransformer\textsuperscript{b, e} \cite{phan2022sleeptransformer} & 21  & spec & Transformer &  0.743 \\ 
		
\hline
       
	\end{tabular}

	\vspace{2em} 
	\captionsetup{justification=centering, singlelinecheck=false, position=below}
\parbox{0.95\linewidth}{\small a: based on SEDF-SC (39 recordings). b: based on SEDF-SC (153 recordings). c: based on full SEDF (197 recordings). d: 20-fold crossvalidation. e:  10-fold crossvalidation. f: 8:1:1 training-validation-test holdout-set evaluation. g: leave-one-subject-out cross validation. *: Training from scratch for 50 epochs.}	
	
\end{table}

\section{Discussion}
 
The S4Sleep(TS) model was shown to lead to competitive performance compared to state-of-the-art methods or even to outperform them \cite{wang2023s4sleep} while only using a moderate input size of 15 epochs. The fact that the model performance did not deteriorate with longer input sizes indicates that the proposed training schedule works as intended. The S4 model as the main building block is particularly known for its ability to capture long-range dependencies across thousands of input tokens, whereas the input size of the predictor module in the S4Sleep(TS) architecture was at most 1000 tokens, even for the case of 200 input epochs. If there were long-range dependencies across hundreds of input tokens that could be exploited diagnostically, the proposed model architecture should have been able to turn them into measurable performance improvements. Therefore the results of our investigation do not support the hypothesis of a diagnostic importance of very long-range interactions for sleep staging.

These results do not contradict the results from \cite{phan2023lseqsleepnet} as the improvements through long-range interactions clearly depend on the model architecture. Their LSTM-based architecture operating on spectrograms as input representations is very different from the S4-based S4Sleep(TS) based on raw time series. However, it is also worth stressing that the S4Sleep(TS) model outperforms the L-SeqSleepNet result for 200 epoch already for 10-15 input epochs and across multiple datasets, see \cite{wang2023s4sleep}. This suggests that the S4Sleep(TS) is able to extract more discriminative features already from smaller input sizes, whereas L-SeqSleepNet needs longer input sequences to (partially) compensate for that.
In our study, our primary aim was to investigate the performance of the best-performing S4Sleep(TS) in the limit of very long input sequences. The fact that no performance improvements where found does not preclude possible performance improvements with models of larger capacity or different model architectures.

\section{Conclusion}
In this study, we revisited the recently proposed S4Sleep(TS) model, a sleep staging model based on structured state space sequence models as central architectural component that operates on raw time series as input. This model was shown to outperform state-of-the-art algorithms on large-scale sleep staging datasets such as SHHS \cite{wang2023s4sleep}. This competitive model performance together with the capabilities of the S4-layer to capture long-range dependencies allowed us to address the question of the importance of very long-range interactions across hundreds of input epochs that was recently put forward in the literature \cite{phan2023lseqsleepnet}. Our results with a careful upscaling of the input size to avoid optimizational issues lead to a model performance that does not change in a statistically significant manner upon increasing the input size. These results put into question the diagnostic value of very long-range interactions for sleep staging.

\ifCLASSOPTIONcaptionsoff
  \newpage
\fi

\bibliographystyle{IEEEtran}
\bibliography{bibfile}

\end{document}